# Title: Speciation: Goldschmidt's Heresy, Once Supported by Gould and Dawkins, is Again Reinstated

**Author:** Donald R. Forsdyke

**Affiliations:** Department of Biomedical and Molecular Sciences, Queen's University, Kingston, Ontario, Canada K7L3N6

Correspondence to: forsdyke@queensu.ca

**Abstract**: The view that the initiation of branching into two sympatric species may not require natural selection emerged in Victorian times (Fleeming Jenkin, George Romanes, William Bateson). In the 1980s paleontologist Steven Jay Gould gave a theoretical underpinning of this non-genic "chromosomal" view, thus reinstating Richard Goldschmidt's "heresy" of the 1930s. From modelling studies with computer-generated "biomorphs," zoologist Richard Dawkins also affirmed Goldschmidt, proclaiming the "evolution of evolvability." However, in the 1990s, while Gould and Dawkins were recanting, bioinformatic, biochemical and cytological studies were providing a deeper underpinning. In 2001 this came under attack from leaders in the field who favored Dawkins' genic emphasis. Now, with growing evidence from multiple sources, we can reinstate again Goldschmidt's view and clarify its nineteenth century roots.

**One Sentence Summary:** Twentieth century evolutionary disputes coalesced around Gould and Dawkins who both supported and then rejected Goldschmidt's now favored chromosomal view.

**Main Text:** There are two groups of hypotheses on the initiation of a branching process that can lead to new species. One group emphasizes the role of natural selection in changing gene frequencies so that species arrival is fundamentally no different than species survival. Macroevolution is merely extrapolated microevolution. It is challenged by a group that downplays the role of natural selection and posits non-genic discordance between members of a species. Macroevolution is not extrapolated microevolution. For some speciations one group may apply. For other speciations the other group may apply. Agreement is sought as to which initiation mechanisms are actually, rather than hypothetically, capable of originating species, and which are most likely to have operated in the general case (*1-3*).

Both groups of hypothesis agree that the "spark" that initiates involves a mechanism for securing reproductive isolation so the process is not subverted by recombination between the genomes of diverging types – such recombination would tend to homogenize rather than retain differences. Both groups also agree that reproductive isolation begins with interruption of the reproductive cycle – gamete, zygote, embryo, meiotic adult gonad, gamete, etc. . Being a recursive cycle, any point, be it before or after union of gametes to form a zygote, will serve to mediate the primary interruption. Many points are susceptible to genic influence. Thus, there may be discordances between paternal and maternal genes affecting gamete transfer or union (prezygotic isolation), or between paternal and maternal genes affecting somatic development or gametogenesis (postzygotic isolation). The gametogenic point is under both genic and non-genic influence. Whatever the point, for successful branching evolution, two independent cycles – two species – must eventually emerge.

While such hypotheses that unify a range of apparently disparate observations are valuable, they should be testable by experimentation and/or by computer modelling. Most speciation events occurred millions of years ago and seem beyond the range of experiment. However, organisms with short generational cycles (viruses, bacteria) show promise in this respect (*4-5*). Furthermore, there are computer simulations with various artificial life models (*6*). Indeed, it will be shown here that one computer model (*7*), accords well with long-held non-genic hypotheses for species initiation.

**Darwin and Jenkin**

The conceptually simplest form of primary reproductive isolation is the separation of members of a population into two types by a geographical barrier. This prezygotic isolation prevents the gametes on one side of the barrier uniting with gametes on the other. Thus, two independently breeding types arise (allopatric speciation). If their isolation is sustained, then other prezygotic and/or postzygotic differences will accumulate so that, should the primary barrier be removed, these secondary differences will then serve as barriers to maintain the reproductive isolation. Under an isolation shield, phenotypic differences appearing in members of branching species will positively or negatively affect their reproductive success in response to natural selection pressures.

This view corresponds closely to that of Charles Darwin (*8*). However, in 1867 Darwin was challenged by Fleeming Jenkin (*9*), who pointed out that adaptive responses to selection pressures must be balanced. Adaptation for flight, for example, invokes counter-adaptations in body weight. Better cognition seems to require an increase in brain weight, but flight imposes a limit on brain size. Thus, one adaptation invokes counter-adaptations, so achieving an organism best balanced to 'pursue' a particular evolutionary strategy. Jenkin saw members of a biological species as enclosed within a sphere that limited their variation. They could evolve so far, but no further. To escape beyond the limits of the sphere – to increase the *evolvability* of the species – *something more* was needed. Thus, there had to be what, in modern terminology can be called an "evolution of evolvability." Once a Jenkinian limit was overcome, the pace of evolution might increase and then slow as a new limit was approached. Thus its rate on a geological time scale might appear "punctuated."

**Evolution of evolvability**

Jenkin, a Scottish professor of engineering with little biological expertise, claimed to be "an impartial looker-on" who would "admit the facts, and examine the reasoning." A century later, as *in silico* modelling of "artificial life" became increasingly feasible, an Oxford professor with considerable biological expertize, Richard Dawkins, began from "what started out as an educational exercise" in 1986 to follow the evolution of computer-generated "biomorphs" (*10*). The "surprising consequences" (*7*) were consistent with Jenkin's viewpoint, although Jenkin was not cited in this respect.

While certain computer programs encoded "embryologies" (developmental plans) that were able cumulatively, under the selection pressure of the human eye, to generate elaborate model organisms, there were found to be limits. When exploring production of a "biomorph alphabet" with his "Blind Watchmaker" computer program, in 1988 Dawkins was "astonished and delighted" to find (*7*):

> There are some shapes that certain kinds of embryology seem incapable of growing. My present *Blind Watchmaker* embryology, that is the basic nine genes plus segmentation with gradients and symmetry mutations, is, I conjecture, forever barred from breeding a respectable K, or a capital B.

4Likewise, when in 1996 he tried to computer-breed radially symmetric starfish, Dawkins observed (*11*) that: "Computer biomorphs can look superficially like echinoderms, but they never achieve that elusive five-way symmetry." Indeed, "the program itself would have to be rewritten for that." In other words, *something more* was needed beyond the selection he was able to impose on a biomorph evolving within the limits that were intrinsic to a given computer program. Dawkins concluded (*7*):

> Huge vistas of evolutionary possibility, in real life as well as in artificial life, may be kept waiting a very long time, if not indefinitely, for a major, reforming change in embryology. … [As for] the evolution of evolvability, … certain kinds of embryology find it difficult to generate certain kinds of biomorphs; other kinds of embryology find it easy to do so. It is clear that we have here a powerful analogy for something important in real biology, a major principle of life that is illustrated by artificial life. It is less clear which of several possible principles it is!

**Gonadal location of a primary isolation mechanism**
A location for a "major principle of life" that might allow escape beyond the limits of Jenkin's sphere, was inferred in 1871 by the London physician St. George Mivart (*12*):

> Now the new forms must be produced by changes taking place in organisms in, after, or before their birth, either in their embryonic, or toward or in their adult, condition. … It seems probable therefore that new species may arise from some constitutional affection of parental forms – an affection mainly, if not exclusively, of their generative system. Mr. Darwin has carefully collected numerous instances to show how excessively sensitive to various influences this system is.

Likewise, Darwin's young research associate, George Romanes (*13*), inferred a constitutional affection of the "reproductive system" where cryptic "collective variations" might accumulate in a sector of a species (see below). This theme was extended by geneticist William Bateson, who was cognizant of Michael Guyer's elegant studies of meiotic chromosomes (*14-17*). A more specific localization to chromosomes of reproductive system germ cells was later postulated by the Danish "father of yeast genetics," Öjvind Winge (*18*).

In 1917 Winge (*19*) interpreted the precise pairing of homologous chromosomes at meiosis in the gonads of diploid organisms as an error-correcting mechanism that demanded close sequence complementarity between parental chromosomes. Should the parental chromosomes have diverged beyond a certain limit – a point of no return – then pairing would fail and their children (hybrids), while appearing phenotypically normal, would be sterile. *Their* sterility was a *parental* phenotype. As manifest in their offspring, the parents were reproductively isolated from each other. Should they remarry, then, with appropriate mates, fertile offspring might be produced. The *couple-specific* defect was a *general* property of the chromosomes themselves and the prediction was made that an experimental genome duplication to generate tetraploid hybrids would "cure" the sterility. Each parental chromosome would then be able to pair with its like at meiosis, as is now recognized (*20*).

**Chromosomes as "reaction systems"**

At that time Goodspeed & Clausen, from studies of crosses between allied species of tobacco plants, were pointing to a *higher order* of organization that lay much above that of individual Mendelian genes (*21*):

> If, for example, it is possible to obtain hybrids involving not a contrast between factors [genes] within a single system, but a contrast of systems all along the line, then it is obvious that we must consider the phenomenon on a higher plane, we must lift our point of consideration as it were from the units of the system [genes] to the systems as units in themselves.

The latter unit systems were referred to as "reaction systems." These seemed to correspond to chromosomes or large parts of chromosomes, and the degree of sterility of offspring correlated positively with differences between such systems:

> When distinct reaction systems are involved, as in species crosses, the phenomena must be viewed in the light of a contrast between systems rather than between specific factor [genic] differences, and the results obtained will depend upon the degree of mutual incompatibility displayed between the specific elements of the two systems. Sterility in such cases depends upon non-specific [non-genic] incompatibility displayed between elements of the system involved, and the degree of this sterility depends upon the degree of such incompatibility rather than upon a certain number of factors concerned in the expression of such behavior.

**Crowther and Bateson**

With prompting from Plymouth physician C. R. Crowther, William Bateson took this further (*22*). In remarks at the 1922 Toronto meeting of the American Association for the Advancement of Science he had attacked the Darwinian notion of species arising as a mere "summation of variations" affecting the conventional phenotype (i.e. microevolution):

> But that particular and essential bit of the theory of evolution which is concerned with the origin and nature of *species* [Bateson's italics] remains utterly mysterious. We no longer feel … that the process of variation, now contemporaneously occurring, is the beginning of a work which needs merely the element of time for its completion; for even time cannot complete that which has not yet begun. The conclusion in which we were brought up, that species are a product of a summation of variations ignored the chief attribute of species first pointed out by John Ray that the product of their crosses is frequently sterile in greater or less degree. Huxley, very early in the debate pointed out this grave defect in the evidence, but before breeding researches had been made on a large scale no one felt the objection to be serious.

Crowther began by noting that, while parental chromosomes had to cooperate for *development* of the zygote from embryo to adult, a far *higher* degree of cooperation would be needed when the chromosomes paired ("conjugated") in the gonad of that adult (*23*):

> Homologous chromosomes … have to cooperate to produce the somatic cell of the hybrid, and their co-operation [for embryo development] might be expected to require a certain resemblance; but for the production of sexual cells [gametogenesis] they must do more, they must conjugate [pair]; and for conjugation it is surely reasonable to suppose that a much more intimate resemblance would be needed. We might, therefore, expect, on purely theoretical grounds, that as species and genera gradually diverged, it would be increasingly difficult to breed a hybrid between them; but that, even while a hybrid could still be produced, a fertile hybrid would be difficult or impossible, since the cells of the germ-track would fail to surmount the meiotic reduction stage when the homologous chromosomes conjugate. This is exactly what happens: the cells go to pieces in the meiotic phase.

Bateson's disparagement of the idea that species might be "a product of a summation of variations" left Crowther "frankly puzzled," for "the proposition is certainly not self-evident." Surely, if the sterility of an offspring were due to a failure within that offspring of homologous chromosomes to pair, it mattered little whether the lack of complementarity responsible for that failure was produced by one large variation, or by the summation of many smaller variations. That Crowther was thinking of primary variations occurring at the chromosomal level, rather than anatomical variations of the sterile individual, was explicit:

> If a sword and scabbard are bent in different directions, it will happen sooner or later that the sword cannot be inserted, and the result will be the same whether the bending be effected by a single blow, or whether it be, in Dr. Bateson's words, 'a product of a summation of variations.' Is this illustration apt? The sword and the scabbard are the homologous chromosomes. … It seems easier to imagine sterility arising from a gradual modification, spread over a length of time, and involving many chromosomes.

Bateson conceded that discontinuity of variation was not critical (*24*):

> It is … not difficult to 'imagine' interspecific sterility produced by a gradual (or sudden) modification. That sterility might quite reasonably be supposed to be due to the inability of certain chromosomes to conjugate, and Mr. Crowther's simile of the sword and the scabbard may serve to depict the sort of thing we might expect to happen.

Thus, Bateson agreed with Crowther that a fundamental form of reproductive isolation, manifest as the hybrid sterility seen when members of allied species were crossed, could be due to an incompatibility characterized cytologically as defective pairing of paternal and maternal chromosomes at meiosis. It was inferred that if we can understand what makes chromosomes incompatible, then we can understand hybrid sterility. And if we can understand hybrid sterility, we can understand an origin of species.

**A modern interpretation**

But how do chromosomes that are homologous (i.e. are *alike*) pair with each other? Do they pair by virtue of this likeness (like-with-like), of by virtue of some key-in-lock (sword-in-scabbard) complementarity, which implies that they are not really alike? One must be the sword and the other the scabbard.

We now appreciate that this paradox was resolved when it was discovered that hereditary information was stored and transmitted as duplex DNA, with two strands – a 'Watson' strand and a 'Crick' strand – that paired with each other by virtue of base complementarity. So, in Crowther's terminology, potentially the sword strand of one chromosome can pair with the scabbard strand of the homologous chromosome (and vice versa). For this swords have to be unsheathed from their own scabbards and then each inserted into the scabbards of the other. Thus the Watson strand of one chromosome must pair with the Crick strand of the other, and vice versa. This requires that the Watson strand be displaced from pairing with the Crick strand of its own chromosome. Likewise, the Crick strand of the homologous chromosome must be displaced from pairing with the Watson strand of its own chromosome. Then cross-pairing can occur.

The pairing requires complementarity of DNA base sequences. A sporadically appearing change in an *individual* base could, if dominant, introduce a new phenotype, but would not greatly affect the overall complementarity between parental chromosomes. However, over time, base changes – including some affecting genes, but *many* not affecting genes – could accumulate. Romanes' "collective variation" that would build up in "a section of a species" (*13-14*) can now be interpreted as a general variation between paternal and maternal DNA sequences. When such differences between chromosomal homologs reached a critical value, meiotic pairing in an offspring's gonads would be impaired and gametogenesis would begin to fail. This early gametogenesis barrier would eventually yield to the developmental and transmission barriers – both of genic origin (*15, 25*). To understand how this view came about, we must go back to the 1930s prior to any appreciation of the structure and role of DNA.

**Chromosomal repatterning**

The idea of chromosomes as "reaction systems" was taken up by Richard Goldschmidt in the 1930s (*26*). Such "reaction systems," through a "repatterning" involving "systemic mutations," might change into other "reaction systems." Thus (*27*):

> The classical theory of the gene and its mutations did not leave room for any other method of evolution. Certainly a pattern change within the serial structure of the chromosome, unaccompanied by gene mutation or loss, could have no effect whatsoever upon the hereditary type and therefore could have no significance for evolution. But now pattern changes are facts of such widespread and, as it seems, typical occurrence that we must take a definite stand regarding their significance. … The pattern changes are in themselves effective in changing the genotype without any change of individual genes. … Point mutations have never been known to change the point-to-point attractions between the homologous chromosomes in the heterozygote. …A repatterning of a chromosome may have exactly the same effect as an accumulation of mutations. … The change from species to species is not a change involving more and more additional atomistic changes, but a complete change in primary pattern or reaction system into a new one, which afterwards may again produce intraspecific variation by micromutation.

Chromosomal "repatterning," namely a change in "the arrangement of the serial chemical constituents of the chromosomes," proceeded slowly and progressively, without necessarily producing any change in the structure or function of organisms, until a new species emerged that was reproductively isolated from the old one by virtue of the new pattern being "incompatible" with that of the old:

> A systemic mutation (or series of such) … consists of a change of intrachromosomal pattern... . Whatever genes or gene mutations might be, they do not enter this picture at all. Only the arrangement of the serial chemical constituents of the chromosomes into a new, spatially different order; i.e. a new chromosomal pattern, is involved. The new pattern seems to emerge slowly in a series of consecutive steps ... . These steps may be without a visible effect until the repatterning of the chromosome ... leads to a new stable pattern, that is, a new chemical system. This may have attained a threshold of action beyond which the physiological reaction system of development, controlled by the new genetic pattern, is so basically changed that a new phenotype emerges, the new species, separated from the old one by a bridgeless gap and an incompatible intrachromosomal pattern.

By "incompatible" Goldschmidt was here referring to differences between the chromosomes of two potential parents. These chromosomes would consequently not be able to cooperate functionally and/or to pair properly at meiosis within their child.

An unlimited number of patterns is available without a single qualitative chemical change in the chromosomal material, not to speak of a further unlimited number after qualitative changes (model: addition of a new amino acid into the pattern of a protein molecule). ... These pattern changes may be an accident, without any significance except for creating new conditions of genetic isolation by chromosomal incompatibility... .

This may seem labored, but in the 1930s it was known neither what genes were chemically, nor how that chemistry might be altered when mutations occurred. However, in the 1870s Ewald Hering and Samuel Butler had laid a framework for thinking about heredity in informational terms (*28*). Striving to give some meaning to his concept of pattern, in 1940 Goldschmidt wrote (*27*):

Let us compare the chromosome with its serial order to a long printed sentence made up of hundreds of letters of which only twenty-five different ones exist. In reading the sentence a misprint of one letter here and there will not change the sense of the sentence; even the misprint of a whole word (rose for sore) will hardly impress the reader. But the compositor must arrange the same set of type into a completely different sentence with a completely new meaning, and this in a great many different ways, depending upon the number of permutating letters and the complexity of the language (the latter acting as a 'selection'). To elevate such a model to the level of a biological theory we have, or course, to restate it in chemical terms.

Four years before Oswald Avery showed that DNA was the form in which hereditary information was transferred through the generations, and thirteen years before Watson and Crick presented a model for DNA, it was not unreasonable to think of chromosomal patterns in terms of amino acid, rather than nucleotide, sequences. Thus Goldschmidt wrote:

> I do not think that an actual chemical model can yet be found. But we might indicate the type of such a model which fulfills at least some, though not all, of the requirements. It is not meant as a hypothesis of chemical chromosome structure, but only as a chemical model for visualizing the actual meaning of a repatterning process … . Let us compare the chromosome to a very long chain molecule of a protein. The linear pattern of the chromosome is then the typical pattern of the different amino acid residues.

Improvements in staining technologies facilitated chromosome studies in dividing cells and leant further support to views on chromosomal incompatibilities as drivers of speciation (*29, 30*). However, in the latter half of the twentieth century the controversy coalesced around two figures, Dawkins (advocate of natural selection affecting genes) and Stephen Jay Gould (advocate of hierarchical levels of selection involving an agency other than natural selection). This gained wide public attention as is related in popular texts such as *The Evolutionists* (*31*) and *Dawkins vs. Gould* (*32*). The major, albeit transient, support for Goldschmidt's non-genic "macroevolutionary" approach to species initiation came in the 1980s from both the theoretical underpinnings of paleontologist Gould (*33*), and the biomorph studies of zoologist Dawkins (*7*).

**Gould and higher level "species selection"**

In "Is a new and general theory of evolution emerging," Gould in 1980 recapitulated Goldschmidt's case (*33*). Arguing that "macroevolutionary trends [speciations] do not arise from the gradual, adaptive transformation of populations, but usually from a higher-order selection operating upon groups of species," Gould distinguished "species selection" – one of various forms of group selection – from conventional natural selection that acts upon individual organisms. Species selection was achieved by "chromosomal alterations in isolating mechanisms, sometimes called the theory of chromosomal speciation." The conventional Darwinian view was that selection *preceded* isolation. Gould reversed this order:

> But in saltational, chromosomal speciation, reproductive isolation comes first and cannot be considered as an adaptation at all. It is a stochastic event that establishes a species by the technical definition of reproductive isolation. To be sure, the later success of this species in competition may depend upon its subsequent acquisition of adaptations; but the origin itself may be non-adaptive. We can, in fact, reverse the conventional view and argue that speciation, by forming new entities stochastically, provides raw material for selection.

Gould had cited the second volume of Romanes' *Darwin, and After Darwin* (1895), but neither in his 1980 paper, nor subsequently, did Gould refer to Romanes' masterpiece – the posthumously published third volume (*13*). Here a case identical to Gould's had been made. In a theory of "physiological selection" Romanes had declared that "diversification of character can never be *originated* by natural selection." A "morphological divergence" guided by natural selection could only be secondary. There was some "physiological peculiarity" of the reproductive system, the basis for which it was for the future to determine. Whatever its basis:

> At least in a large number of cases, it was the physiological peculiarity which first of all led to the morphological divergence, by interposing the bar of sterility between two sections of a previously uniform species; and by thus isolating the two sections one from another, started each upon a subsequent independent course of divergent evolution. … In the absence of other forms of isolation [e.g. geographical], the morphological divergences could not have taken place at all, had not the physiological peculiarity arisen.

Praising Goldschmidt for having provided a logical basis for "species selection," in 1982 Gould encouraged, and wrote a forward for, a reprinting of Goldschmidt's *The Material Basis of Evolution* (*34*). However, Gould rejected Goldschmidt's notion of "'systemic mutations' involving the entire genome." On the other hand, Romanes (*13*) had characterized his "physiological peculiarity" as "a 'collective variation' affecting a number of individuals simultaneously, and therefore characterizing a whole race or strain" (i.e. a section of a species). This is consistent with our modern understanding. The Goldschmidtian abstractions can now be fleshed out in both bioinformatic and molecular terms (*15, 25, 26*).

**Dawkins and higher level "species selection"**

Apparently overlooked by Gould, powerful support for Goldschmidt came in 1988 from the computer simulations of Dawkins (*7*). Like Goldschmidt's microevolution/macroevolution dichotomy, as a result of his biomorph simulations Dawkins called for a distinction "between two kinds of mutation: ordinary changes within an existing genetic system, and change to the genetic system itself." The former were "the standard mutations that may or may not be selected in normal evolution within a species." The latter were "changes to genetic systems [that] must have been, at least in one sense, major changes, changes of a different order from the normal allele substitutions that go on within a genetic system." The latter class of mutation were associated with "changes in embryology which … are … evolutionarily pregnant." Thus:

> As the ages go by, changes in embryology that increase evolutionary richness tend to be self-perpetuating. … I am talking about a kind of higher-level selection, a selection not for survivability but for evolvability. … Others have pointed out that we should speak of 'species selection' only in those rare cases where a true species-level quality is being evolved. Species selection, for instance, should not be invoked to explain an evolutionary lengthening of the leg, since species don't have legs, individuals do. It might, on the other hand, be invoked to explain the evolution of a tendency to speciate, since speciating is a thing species, but not individuals, do. It now seems to me that an embryology that is pregnant with evolutionary potential is a good candidate for a higher-level property of just the kind that we must have before we allow ourselves to speak of species or higher-level selection.

However, unlike Gould whose higher-level selection was due to "a stochastic event," Dawkins would not stray from natural selection and did not see a relationship between his own work and that of Goldschmidt (*7*):

> Perhaps there is a sense in which a form of natural selection favors, not just adaptively successful phenotypes, but a tendency to evolve … . I have been in the habit of disparaging the idea of 'species selection' … . But selection among embryologies for the property of evolvability, it seems to me, may have the necessarily qualifications to become cumulative in evolutionarily interesting ways. After a given innovation in embryology has been selected for its evolutionary pregnancy, it provides a climate for new innovations in embryology. Obviously the idea of each new adaptation serving as the background for the evolution of subsequent adaptations is commonplace, and is the essence of the idea of cumulative selection. What I am now suggesting is that the same principle may apply to the evolution of evolvability, which, therefore, may also be cumulative.

**Uncoupling of speciation from adaptation**

Sadly, in *The Structure of Evolutionary Theory* (2002) Gould recanted his Goldschmidtian heresy of the 1980s, while still maintaining "a hierarchical theory of selection." Thus (*35*):

> I do not, in fact or retrospect … regard this 1980 paper as among the strongest … that I have ever written … . I then read the literature on speciation as beginning to favor sympatric alternatives to allopatric orthodoxies at substantial relative frequency, and I predicted that views on this subject would change substantially, particularly towards favoring mechanisms that would be regarded as rapid even in microevolutionary time. I now believe I was wrong in this prediction.

Likewise we peruse in vain *Brief Candle in the Dark* (2015) for an expanded recognition of the implications of Dawkins' biomorph studies (*36*). Forgetting what was once "drummed into my innermost consciousness" (*7*), Dawkins remains a self-proclaimed "dyed-in-the-wool, radical neoDarwinian" reiterating previous condemnations of the "utter nonsense" perpetrated by William Bateson. A full case for reinstallation of the Goldschmidtian certitudes, once so lavishly entertained and then later disavowed by these great evolutionists, is set out elsewhere (*25*). Here, a few quotations must suffice.

In 2007 Greig found that "speciation genes do not play a major role in yeast speciation," and proposed that "simple sequence divergence is the major cause of sterility in F1 hybrids formed between *S. cerevisiae* and *S. paradoxus*" (*37*). Commenting on this Louis (*38*) concluded that "one must be cautious in labelling gene incompatibilities as speciation genes, or at least in interpreting them as being causal in the speciation process rather than a result of divergence post-speciation."

In 2010 Venditti *et al*. noted that a dependence of branching on *synonymous* mutations, which do not change the encoded amino acid, seemed to exclude natural selection as a general initiator of species divergence and linked "speciation to rare stochastic events that cause reproductive isolation." Thus (*39*):

> Species do not so much 'run in place' as simply wait for the next sufficient cause of speciation to occur. Speciation is freed from the gradual tug of natural selection; there need not be an 'arms race' between the species and its environment, nor even any biotic effects. To the extent that this view is correct, the gradual genetic and other changes that normally accompany speciation may often be *consequential* to the event that promotes the reproductive isolation, rather than causal themselves.

Likewise, Hedges *et al*. (*40*) from analyses of 50,000 eukaryotic species, infer "an uncoupling of speciation from adaptation," and conclude that "adaptive change that characterizes the phenotypic diversity of life would appear to be a separate process from speciation." Furthermore, Bhattacharyya *et al*. from mouse breeding studies point to non-genic (non-coding) sequence differences as a basis for hybrid sterility (*41*):

> We propose the heterospecific pairing of homologous chromosomes as a preexisting condition of asynapsis [failure of chromosome pairing] in interspecific hybrids. The asynapsis may represent a universal mechanistic basis of F1 hybrid sterility manifest as pachytene arrest. It is tempting to speculate that a fast-evolving subset of the noncoding genomic sequence important for chromosome pairing and synapsis may be the culprit.

As for molecular footprints of isolation mechanisms, in 2013 Lawrie *et al*. (*42*) reported strong selection at *synonymous* sites in fruit fly and concluded that: "The underlying biological function disrupted by these [synonymous] mutations is unknown, but it is not related to the forces generally believed to be the principal actors shaping the evolution of synonymous sites."

That such a force might relate to speciation and DNA base composition ("1-mer" base frequencies; *43*) is generalizable to higher oligonucleotide compositions ("k-mer" frequencies; *25*). Consistent with this, Brbic *et al*. (*44*) note that there is a conflict between the needs of DNA, and of the proteins it encodes, that strongly favors DNA. Genomes are *dominated* by oligonucleotide frequencies that can *overrule* the needs of protein-encoding to the extent that the latter is reflected in amino acid compositions:

> We find that G + C content, the most frequently used measure of genomic composition, cannot capture diversity in amino acid compositions and across ecological contexts. However, di-/trinucleotide composition in intergenic DNA predicts amino acid frequencies of proteomes to the point where very little cross-species variability remains unexplained. … A corollary is that the previously proposed adaptations of proteomes to environmental challenges … may need to be reinterpreted, while taking into account the evolutionary forces shaping DNA oligonucleotide frequencies.

**Conclusion**

Having reached a degree of consensus on speciation in the 1980s, two influential "public intellectuals" later declined to reinstate it. Discounting the historical and growing new evidence for the consensus, there developed a notorious dispute between these celebrity scientists much of which, with other perhaps prejudicial views (1, 3), must now be laid to rest.

**Acknowledgments:** Queen's University hosts my web-pages where much of the early literature, including the relatively inaccessible texts of Guyer and Winge, has been made available.